\documentclass[a4paper, 11pt]{article}
\usepackage{graphicx}
\usepackage{caption}
\usepackage{rotating}
\usepackage{geometry}
\usepackage{hyperref}
\begin{document}
\begin{center}
{\Large\textbf{$H_0$ Tension: Response to Riess et al arXiv:1810.03526}}

Tom Shanks, Lucy Hogarth and Nigel Metcalfe

Department of Physics, Durham University, South Road, Durham DH1 3LE, England

\end{center}
\medskip

\noindent{\bf \underline {Abstract}}
Riess et al (2018c, R18c) have claimed there exist seven problems in the
analyses presented by Shanks et al (2018, S18) where we argue that there
is enough uncertainty in Cepheid distances and local peculiar velocity
fields to explain the current  tension in $H_0$. Here, we take each of
the R18c points in turn and suggest that either they do not apply or
that the necessary caveats are already made by S18. We conclude
that the main point to be inferred from our analyses still stands which
is that previous claims by Riess et al (2018b) that Gaia parallaxes confirm
their Cepheid scale are, at best, premature  in advance of further 
improvements in the Gaia astrometric solution.

\section{Gaia parallaxes and the Cepheid scale}

\noindent\underline {1. The inclusion of Cepheids in open clusters.} 
R18c criticise our inclusion of Cepheids whose distances were estimated by main-sequence 
fitting open clusters of which they are members because ``These distances have never been 
used in modern $H_0$ determination, and their values are thus irrelevant in the context of 
distance scale.'' It is true that they are not used in the distance scale of Riess et al (2016, 2018b)
but they have played a  crucial role in the distance scale for the past $\approx60$ years.
(see eg Irwin 1955; Feast 1957; Feast \& Walker 1987; Laney \& Stobie 1993
Feast 2003; Hoyle, Shanks \& Tanvir 2003; Sandage \& Tammann 2006; An et al 2007;
Turner et al 2008; Anderson, Eyer and Mowlavi 2013; Chen, de Grijs \& Den 2017; Lohr et al 2018).
Indeed, it was noticing that the Gaia estimates of  these open cluster/Cepheid  
distances were significantly longer than previously estimated that prompted the start of our study.
We clearly state  in our paper that these open cluster Cepheids give `contextual evidence
that the Gaia distances imply a longer Cepheid scale', while looking to another route 
to provide the most direct comparison with the scale of Riess et al (2016, 2018b).

\smallskip

\noindent \underline{2. The inclusion of Cepheids with saturated Gaia
magnitudes.} R18c criticise our comparison of the HST FGS sample of
Benedict et al (2007) due to their having saturated magnitudes with
$G<6$, saying  ``the Gaia Team strongly recommends parallaxes for
saturated objects in DR2 not be used, calling them `unreliable'
(Lindegren et al. 2018)''. In fact, Lindegren et al make the less strong
statement that ``The bright limit is $G\approx3$, although stars with
$G<6$ generally have inferior astrometry due to calibration issues.''
Moreover, in their selection of an astrometrically clean $<100$pc sample
Lindegren et al also  include stars in the $3<G<6$ range (see their Fig. C2
(left)). We also excluded the 3 stars with the lowest parallax/error
ratio and the remaining 5 all had distances within 600pc. Nevertheless,
we acknowledge that stars with $G<6$ may still have systematic
uncertainties and Riess et al (2018b) take this view, arguing from their Fig. 4 that there
is a much larger Gaia offset from their parallaxes at brighter $G$ mags
than fainter $G$ mags. However, replotting the data of Fig. 1 of S18
where we now look at corrected Gaia distance modulus minus the previous
modulus versus $G$ mag (see Fig. 1 below), we show  that the {\it
fractional} distance difference is roughly constant with magnitude. This
is consistent with the fractional difference being as large at $G>6$ as
at $G<6$, which could be interpreted as any systematic error due to
saturation being sub-dominant at these low distances and large
parallaxes.

\smallskip

\noindent\underline{3. The Gaia parallax offset.} R18c criticise S18 for
assuming the basic -29$\mu$as quasar offset. In the Riess et al (2018b)
analysis of a sample of 46 Cepheids with distances estimated from their
P-L relation based on  HST photometry and then converted into a
`photometric parallax', they prefer to compare with Gaia parallaxes by
fitting  for a scale error and offset between the two. They find a
result consistent with unit slope (ie no scale error) and an offset of
-46$\mu$as. They interpret this result as the unit slope confirming
their Cepheid scale (and $H_0$) while attributing the offset to Gaia.
Riess et al (2018b) argue that the fact that in their Figs. 5, 6 they
clearly statistically determine that the difference between the Gaia
parallaxes and their own estimates is almost wholly an offset with the
comparison giving almost exactly unit slope, means that their $H_0$ is
confirmed. However, we see no reason for attributing all of the offset
error to Gaia which essentially amounts to calibrating the Gaia offset
on the previous Cepheid scale. Thus as a counter-example, let us now assume
that actually Gaia measures parallaxes perfectly and that the above fit
results of slope and offset still apply. On the argument of Riess
et al the unit slope would argue that their Cepheid scale and $H_0$ is
confirmed. Yet all the perfect Gaia parallaxes would still be a constant
amount smaller than those of Riess et al. This would force the
conclusion that some other problem affected the Riess et al parallaxes
that mimicked a constant  offset. So, for example, even if all their 46
photometric Cepheids had the same period, there could still be  a
problem, perhaps in their de-reddening since more distant Milky Way Cepheids
tend to be more dust absorbed. Clearly, a constant  offset between
parallaxes is suggestive of a Gaia parallax problem but not a proof. This
potential circularity of the Riess et al (2018b) argument is then
repeated in Fig. 1 of R18c where they now argue for a magnitude
dependent Gaia parallax offset solely on the basis of a comparison with
their own Cepheid scale. 

\smallskip

\noindent\underline {4. The error on the Gaia parallax offset.} R18c
criticise S18 for not quoting an uncertainty on our $-29\mu$as Gaia
parallax offset. But here we are following Lindegren et al who treat the
statistical uncertainty on this offset as negligible at $\approx\pm1\mu$as,
given it is based on a standard error from $\approx500000$ quasar
parallax estimates. S18 also  clearly state that there are possible
systematic errors based on magnitude and colour but these are difficult
to quantify at the present time. We then assumed no colour or magnitude
systematics, broadly similar to Riess et al (2018b), and we simply assumed the most
basic quasar offset as representing the first-order DR2 parallax verdict on
the previous Cepheid scale, finding that generally Gaia distances are
longer.

\smallskip

\noindent\underline{5. Other Cepheid calibrations.} R18c note that Riess
et al (2016) base their Cepheid calibration on detached eclipsing
binaries and the water maser in NGC4258 as well as parallaxes. We also
already note these other routes in our paper and say that it will be
interesting to see how they fare as the  Gaia astrometric solution
improves (see Sect. 2 of S18). However, the main focus of S18 is the
current Gaia DR2 parallaxes and the level to which they support the
Cepheid scale of Riess et al (2016, 2018b). In their paper and elsewhere
(see \url{https://www.nasa.gov/feature/goddard/2018/hubble-and-gaia-
team-up-to- fuel-cosmic-conundrum}), Riess et al (2018b)  claim that the
Gaia DR2 parallaxes confirm their previous Cepheid scale. We have argued
above against this view and conclude that either the DR2 parallaxes are
currently too uncertain for absolute Cepheid calibration or in their
simplest form they give evidence against the previous scale. Thus
deciding whether there is any  inconsistency between parallax, the
other two Cepheid calibration routes and the Cepheid scale in general
must await improved Gaia astrometry.

\section{The Local Hole}

\noindent\underline {6. The Local Hole and the fit of the SNIa Hubble
diagram.} Here R18c claims that the SNIa Hubble diagram fit of S18 is
worse with the Local Hole outflow effect included. But here R18c only
compares to our fit with $\Omega_m$ kept fixed at its original value of
$\Omega_m=0.28$. When S18 fitted for both $H_0$ and $\Omega_m$ we found
that including the effect of the Local Hole actually marginally improved
the fit, reducing the total  $\chi^2$  from 1035.7 to 1034.1. The fitted
value of $\Omega_m$ only rose from 0.28 to 0.33 compared to the Planck
value of 0.31 so we conclude the Local Hole outflow is consistent  with
the SNIa Hubble diagram.

\smallskip

\noindent\underline{7. Cosmic variance of Local Hole.} Here R18c claim
that our Local Hole model implies a $6\sigma$ fluctuation in a
$\Lambda$CDM model based on the variances in simulations studied by Wu
\& Huterer (2017) and Odderskov et al (2017), quoting an estimated 0.3\%
fluctuation in $H_0$. But Wu \& Huterer report a standard deviation of
0.31 kms$^{-1}$Mpc$^{-1}$, a 0.46\% fluctuation which makes our 1.8\%
rise a $3.9\sigma$ fluctuation rather than $6\sigma$. Odderskov et al
emphasise the differences that different SNIa selections make quoting
fluctuations in the range  0.3-0.96\% for two previous SNIa surveys, so
by the latter estimate  the Local Hole significance could be as low as $\approx1.9\sigma$.
But note that we have previously accepted that if the Local Hole extends over a
large fraction of the sky, it may start to challenge the standard cosmology 
at a similar $2.5-4\sigma$ level (Frith, Metcalfe \& Shanks, 2006).

\section{Conclusions}
We conclude that Gaia DR2 parallaxes do not immediately confirm the
Cepheid scale of  Riess et al (2018b). Indeed, the most direct
interpretation of the new Gaia data is that they support an upwards
revision of previous Cepheid distances by $\approx7$\% in the case of
the Riess et al scale and by up to $\approx18$\% in other cases. We also
conclude that there is strong evidence for a Local Hole that can cause a
further $\approx2$\% reduction in $H_0$. We therefore stand by our
conclusion that Gaia parallaxes, the Local Hole and associated
uncertainties may potentially relieve the previous $H_0$ tension.


\begin{figure}
\begin{center}
    	\includegraphics[height=10cm]{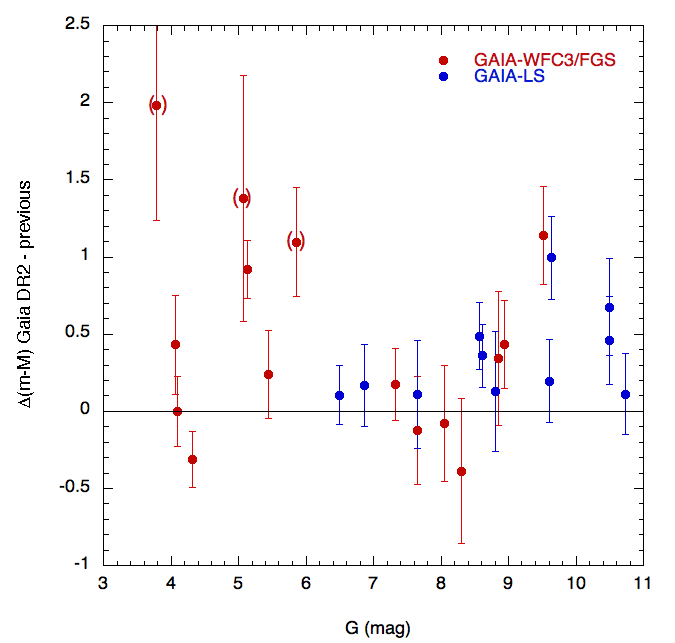}
    	\caption{Data from  S18 Fig. 1 to show how the difference
    	between the $+29\mu$as corrected Gaia distance moduli and the
    	previous distance moduli of Benedict et al (2007, FGS), Riess et
    	al (2016, 2018a, WFC3) and Laney \& Stobie (1993, LS) varies
    	with $G$ magnitude. There is little evidence of  magnitude
    	dependence despite the FGS parallax stars of Benedict el
    	(2007) having saturated $G < 6$ magnitudes. Brackets show the 3
    	stars, $l$ Car, W Sgr and RT Aur of Benedict et al (2007) that,
    	as well as being saturated, have the lowest parallax/error ratio
    	and were  excluded by S18. For the remaining stars,  the corrected Gaia distance moduli are
    	$0.35\pm0.11$mag larger or $17.6\pm5$\% greater in distance.}
\end{center}
\end{figure}

\medskip

\noindent{\bf\underline{References}} 
An D., Terndrup D.~M., Pinsonneault M.~H., 2007, ApJ, 671, 1640
$\bullet$ Anderson R. I., Eyer L., Mowlavi N., 2013, MNRAS, 434, 2238
$\bullet$ Benedict G. F., et al., 2007, AJ, 133, 1810
$\bullet$ Chen X., de Grijs R., Deng L., 2017, MNRAS, 464, 1119 
$\bullet$ Feast M.W., 1957, MNRAS, 117, 193
$\bullet$ Feast M.W., Walker A.R., 1987, ARA\&A, 25, 345
$\bullet$ Feast M.W., 2003, LNP, 635, 45
$\bullet$ Frith W.J., Metcalfe N., Shanks T., 2006, MNRAS, 371, 1601
$\bullet$ Hoyle F., Shanks T., Tanvir N. R., 2003, MNRAS, 345, 269
$\bullet$ Irwin J.B., 1955, MNSSA, 14, 38
$\bullet$ Laney C. D., Stobie R. S., 1993, MNRAS, 263, 921
$\bullet$ Lindegren L., et al., 2018, A\&A, 616, A2
$\bullet$ Lohr M.~E., Negueruela I., Tabernero H.~M., Clark J.~S., Lewis F., Roche P., 2018, MNRAS, 478, 3825 
$\bullet$ Odderskov I., Hannestad S., Brandbyge J., 2017, JCAP, 3, 022 
$\bullet$ Riess A. G., et al., 2016, ApJ, 826, 56 
$\bullet$ Riess A. G., et al., 2018a, ApJ, 855, 136 
$\bullet$ Riess A. G., et al., 2018b, ApJ, 861, 126 
$\bullet$ Riess A. G., et al., 2018c, preprint, (arXiv:1810.03526) (R18c)
$\bullet$ Sandage A., Tammann G.A., 2006, ARA\&A, 44, 93 
$\bullet$ Shanks T., Hogarth L., Metcalfe N., 2018, preprint, (arXiv:1810.02595) (S18)
$\bullet$ Turner D.G., et al., 2008, MNRAS, 388, 444
$\bullet$ Whitbourn J. R., Shanks T., 2014, MNRAS, 437, 2146
$\bullet$ Wu H.-Y., Huterer D., 2017, MNRAS, 471, 4946 

\end{document}